\documentclass[nonacm,sigplan]{acmart}

\usepackage[]{hyperref}

\usepackage{listings}
\definecolor{dkgreen}{rgb}{0,0.6,0} 
\lstdefinestyle{customasm}{
    language=[x86masm]Assembler,
    emph={%
      subs,
      b,
      b.lt,
      b.le,
      b.gt,
      lsr,
      ldr
    },emphstyle={\color{blue}}
}

\usepackage{subcaption}
\usepackage{multirow}

\begin{document}

\title{Dissecting Conditional Branch Predictors of Apple Firestorm and Qualcomm Oryon for Software Optimization and Architectural Analysis}

\author{Jiajie Chen}
\affiliation{%
  \institution{Tsinghua University}
  \country{}
}
\email{cjj21@mails.tsinghua.edu.cn}

\author{Peng Qu}
\affiliation{%
  \institution{Tsinghua University}
  \country{}
}
\email{qp2018@tsinghua.edu.cn}

\author{Youhui Zhang}
\affiliation{%
  \institution{Tsinghua University}
  \country{}
}
\email{zyh02@tsinghua.edu.cn}

\begin{abstract}
  Branch predictor (BP) is a critical component of modern processors, and its accurate modeling is essential for compilers and applications. However, processor vendors have disclosed limited details about their BP implementations.

  Recent advancements in reverse engineering the BP of general-purpose processors have enabled the creation of more accurate BP models. Nonetheless, we have identified critical deficiencies in the existing methods. For instance, they impose strong assumptions on the branch history update function and the index/tag functions of key BP components, limiting their applicability to a broader range of processors, including those from Apple and Qualcomm.

  In this paper, we design a more general branch prediction reverse engineering pipeline that can additionally recover the conditional branch predictors (CBPs) of Apple Firestorm and Qualcomm Oryon microarchitectures, and subsequently build accurate CBP models. Leveraging these models, we uncover two previously undisclosed effects that impair branch prediction accuracy and propose related solutions, resulting in up to 14\% MPKI reduction and 7\% performance improvement in representative applications. Furthermore, we conduct a comprehensive comparison of the known Intel/Apple/Qualcomm CBPs using a unified standalone branch predictor simulator, which facilitates a deeper understanding of CBP behavior.
\end{abstract}

\maketitle 
\pagestyle{plain} 

\section{Introduction}

Modern processors heavily rely on branch prediction to sustain high instruction throughput. To achieve this, the branch predictor (BP) records recent branches and predicts the direction and target address of the current one well before instruction decoding and execution \cite{combining,study}. Given that branch instructions constitute approximately 15\% to 30\% of the total number of executed instructions \cite{reduction}, the design of accurate BPs and software optimization based on accurate BP modeling are of significant importance.

Although the general structure of the branch predictor (BP) is well-known and has been published by some CPU vendors \cite{zen2analysis}, the detailed behavior, including the characteristics of branch history recording, the algorithm for conditional branch prediction, and the capacity of prediction tables (if any), remains largely undocumented. While limited performance monitoring counters (PMCs) can indicate which branches are most frequently mispredicted, programmers may still struggle with optimization, as the BP operates as a black box. In contrast, with an accurate model of the BP of commercial processors, we can simulate the application using the model and pinpoint the actual sources of mispredictions.

Recently, several studies \cite{spectre,halfhalf,trustzonetunnel} have successfully reverse-engineered the conditional branch predictor (CBP) of Intel and ARM Cortex processors. However, our analysis reveals that their reverse engineering pipelines rely on strong assumptions that do not hold for other processors. For instance, the existing method is tailored to the relatively simple PHT index and tag functions of Intel and assumes that the path history register (PHR) can be easily cleared to zero. In contrast, Apple Firestorm (Apple M1 P-core) and Qualcomm Oryon (in Qualcomm X Elite) exhibit much greater complexity in these aspects, rendering the existing method inapplicable.

Therefore, we propose a new and more general pipeline to additionally reverse-engineer the CBPs of Apple Firestorm and Qualcomm Oryon. Specifically, we design new microbenchmarks with minimal assumptions about CBP internals and develop a comprehensive approach to recover all possible combinations of bits in the index/tag functions of all pattern history tables (PHT). Based on this proposal, we recover the novel branch history organization used in these two microarchitectures, as well as the six pattern history tables in their TAGE branch predictor. Furthermore, we reveal the index and tag functions of the PHTs and the set-associative structure of each table. Notably, we identify some hidden PMCs that record conditional branch mispredictions in Oryon, enabling the isolation of CBP from other BP components, which is a prerequisite for the viability of the aforementioned microbenchmarks.

Utilizing the reverse engineered results, we construct an accurate CBP model and use it to locate the root cause of two previously undisclosed effects, named \textbf{Scatter} and \textbf{Annihilation}, which contribute to increased branch misprediction rates. \textbf{Scatter} occurs when one branch maps to an excessive number of PHT entries, while \textbf{Annihilation} refers to the scenario where two or more branches leave identical traces in the branch history. We also devise solutions to mitigate these effects. It is important to note that these phenomena also exist in other general-purpose processors, such as Intel.

Finally, we compare the CBPs of several microarchitectures from Intel, Apple, and Qualcomm by porting them to an in-house standalone branch predictor model and evaluating their branch prediction accuracy on SPEC INT 2017 \cite{spec2017} and Geekbench 5 \cite{geekbench5} benchmarks. Our analysis reveals that capacity significantly impacts MPKI, far more than other microarchitecture-specific functions. We also provide microarchitectural insights on optimizing CBP accuracy and bandwidth that may guide future iterations of branch predictors for both open-source and commercial processors: incorporating additional PC bits into the PHR footprint, increasing the CBP capacity, and partitioning PHTs by one PC bit to enable the prediction of two conditional branches per cycle without a second SRAM read port.
The contributions of this paper are as follows:

\begin{itemize}
  \item For the first time, we recover the CBPs of Apple Firestorm and Qualcomm Oryon. We design a new reverse engineering pipeline that is less restrictive and more general than existing methods.
  \item We construct a CBP model and identify two previously undisclosed effects that impair branch prediction accuracy. Using this model, we identify the root causes of these effects and enhance performance through minimal software/hardware modifications.
  \item We compare the CBPs of commercial processors in terms of their design and MPKI on SPEC INT 2017 and Geekbench 5 benchmarks, providing insights that support future branch prediction development.
\end{itemize}

\section{Background and Related Work}

In this section, we provide an overview of the structure of modern branch predictors and discuss recent advancements in reverse engineering conditional branch predictors.

\subsection{Branch Prediction}

Modern out-of-order processors typically comprise a frontend and a backend. The frontend is responsible for fetching and decoding instructions, while the backend executes and commits them. Branches are prevalent in applications \cite{reduction}, and accurate branch prediction is crucial for ensuring correct instruction fetching and maintaining the continuity of execution flow, which is essential for enhancing program performance.

As a key functional unit, the branch predictor (BP) often includes the following components to predict branch outcomes (direction and target address) early in the frontend: The Branch Target Buffer (BTB) records branch information, including branch type and target address, for branches encountered in recent executions.
The Conditional Branch Predictor (CBP) predicts the direction of conditional branches, which can be either taken (T) or not-taken (NT).
The Indirect Branch Predictor (IBP) predicts the target addresses of indirect branches, which may jump to various addresses.
The Return Address Stack (RAS) predicts the return address of functions.

This paper primarily focuses on the CBP. For the CBP to function effectively, the predictor may utilize local history or global history as input. Local history records the previous directions of the same branch, while global history records the directions or addresses of recent branches. The former correlates a branch with itself, whereas the latter identifies correlations between different branches.

Global history is typically maintained in a Global History Register (GHR) \cite{ghr}, which is a shift register storing the directions of recent branches. For each executed conditional branch, the GHR is shifted by one and updated with the direction (0 for not-taken, 1 for taken) of the last branch. While the GHR scheme is simpler and less costly to implement, it cannot distinguish between different conditional branches or capture correlations between conditional and unconditional branches.

An alternative global history scheme is the Path History Register (PHR) \cite{phr,ogehl,ogehl2}, which is also a shift register. However, instead of storing directions, it records the branch and target address of taken branches. Taken branches typically include conditional branches that are taken, unconditional jumps, indirect jumps, calls and returns. Not-taken conditional branches are not recorded in the PHR. Since storing the entire branch and target address would require significant space, the address bits are typically XOR-ed into a fixed-width footprint, which is then XOR-ed into the shifted PHR. The PHR update function is as follows:

\begin{equation*}
  \mathrm{PHR}_{\mathrm{new}} = (\mathrm{PHR}_{\mathrm{old}} \ll \mathrm{shamt}) \oplus \mathrm{footprint}
\end{equation*}

The definition of branch address varies: in Intel processors, it points to the last byte of the branch instruction, whereas in ARMv8 processors, it points to the first byte. The PHR scheme provides more detailed history but is more expensive to update and recover.

\begin{figure}[htbp]
  \centering
  \includegraphics[width=\linewidth]{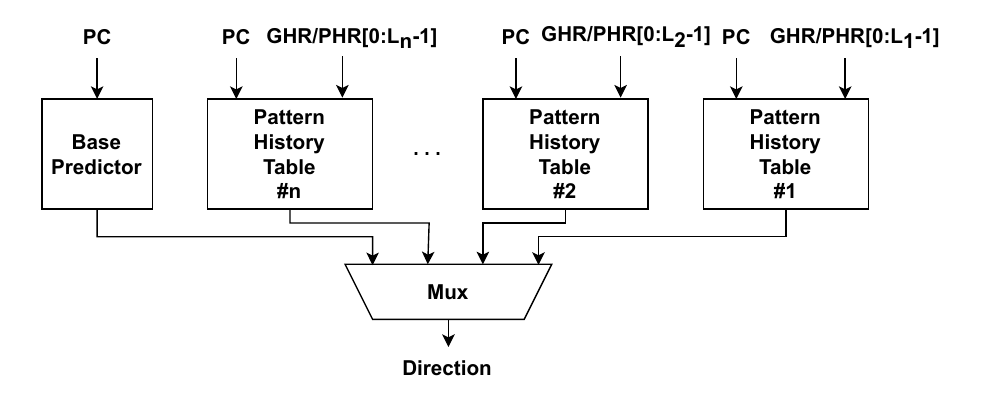}
  \caption{Structure of TAGE conditional branch predictor}
  \label{fig:tage}
\end{figure}

For both global history schemes, the history register and the branch address are used together to predict the direction of conditional branches.
A state-of-the-art conditional branch predictor, TAGE, was proposed by Seznec \cite{tage} in 2006 and has won several branch predictor championships \cite{cbp4,cbp5}. TAGE is widely used in commercial processors, including AMD Zen 2 \cite{zen2analysis}, ARM Cortex-A53 \cite{trustzonetunnel}, and Intel Alder Lake \cite{halfhalf}.
TAGE consists of multiple pattern history tables (PHTs) using different history lengths of the history register and program counter (PC) as inputs. During prediction, the PC and history register are sent to the base predictor and multiple PHTs, and the predicted direction is computed from the matching entries.
Each PHT is a set-associative structure where each entry contains \textit{tag}, \textit{counter}, and \textit{useful} bits. The tag and index are computed from the PC and truncated history register. If there is a tag match, the \textit{counter} value predicts the direction. The history lengths of the PHT inputs $L_1, L_2, \cdots, L_n$ form a geometric series. The structure of the TAGE conditional branch predictor is illustrated in Figure \ref{fig:tage}.

In addition to prediction accuracy, prediction bandwidth is also critical. A simple implementation predicts at most one branch per cycle, making the branch prediction structures easier to implement with low area and high clock frequency. However, many applications execute a large fraction of branches, such as loops with small bodies, where the frontend can deliver at most one loop body's worth of instructions per cycle. Therefore, predicting more branches per cycle is necessary for better performance.

A common method to improve branch prediction bandwidth is to predict two adjacent conditional branches in a cycle. If the first conditional branch is predicted as taken, the branch prediction unit redirects to the target address of the branch. Otherwise, the prediction result of the second conditional branch is examined. Since the two branches are adjacent, it is possible to save these two branches in one BTB entry to be accessed in one cycle \cite{btb,exynos,zen1,zen2,zen3,zen4,zen5}. However, some implementations only support predicting at most one taken branch per cycle \cite{btb,exynos,zen1,zen2,zen3}. AMD Zen 4 and Zen 5 support predicting two taken branches per cycle, further enhancing branch prediction bandwidth \cite{zen4,zen5}.

\subsection{Reverse Engineering Branch Prediction}

In 2019 Google Zero identified the Spectre and Meltdown vulnerabilities in commercial processors which exploit branch target injection \cite{spectre}. To inject arbitrary addresses into the branch predictor, they reverse-engineered the branch predictors in the Intel Haswell microarchitecture. By leveraging the branch prediction internals, it is possible to attack a victim process via branch target injection.

Half\&Half \cite{halfhalf} extended the PHR reverse engineering to more recent Intel microarchitectures, from Haswell to Alder Lake. It revealed that Intel has maintained the same CBP design for many years, primarily increasing history lengths to capture longer branch histories. In addition to the PHR, Half\&Half provided the first detailed implementation of the TAGE CBP in Intel processors, including the set-associative structure of pattern history tables and the index and tag functions used to access the tables. Building on Half\&Half, TrustZoneTunnel \cite{trustzonetunnel} reverse-engineered the CBP of the ARM Cortex-A53 CPU.

The aforementioned work forms the basis for the microbenchmarks used in this paper. However, there are significant differences. For instance, Half\&Half \cite{halfhalf} is designed for the simple design of Intel CBP, which combines branch and target addresses into one footprint and uses only one PC bit in the index function of the PHT, along with typical PHR folding schemes in both index and tag functions.
Additionally, the microbenchmarks in Half\&Half heavily rely on setting the PHR to known values, which is not applicable to Firestorm and Oryon.

Notably, no prior work has revealed the CBPs of the Firestorm and Oryon microarchitectures. Our work addresses this gap by reverse-engineering their CBPs, including the special design of the PHR and the six pattern history tables.

\section{Reverse Engineering CBPs of Apple Firestorm and Qualcomm Oryon}

In this section, we introduce a novel reverse engineering pipeline for conditional branch predictors (CBPs) and apply it to recover the CBPs of Apple Firestorm and Qualcomm Oryon microarchitectures.

\subsection{Reverse Engineering Pipeline}

\begin{figure}[htbp]
  \centering
  \includegraphics[width=0.9\linewidth]{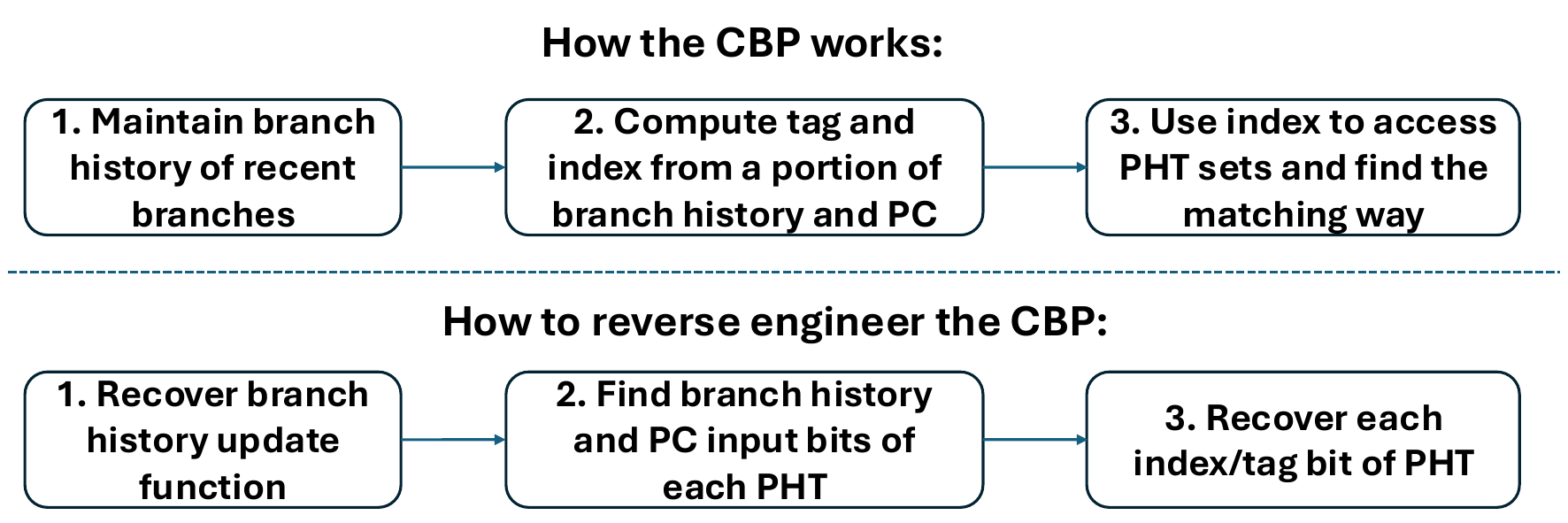}
  \caption{The procedure of CBP and reverse engineering it}
  \label{fig:procedure}
\end{figure}

Before delving into the reverse engineering pipeline, it is essential to revisit how the CBP operates, as illustrated in Figure \ref{fig:procedure}. The CBP records the branch history of recent branches, computes the tag and index based on the branch history and program counter (PC), and finally uses the index to access the pattern history table (PHT) sets and find the matching way. Accordingly, to reverse engineer the CBP, we need to recover the branch history first, then narrow the input range of the PHTs, and eventually recover each index and tag bit of the PHT.

Based on this high-level procedure, existing branch prediction reverse engineering works \cite{halfhalf,indirector} construct a reverse engineering pipeline as shown in Figure \ref{fig:pipeline}. Assuming the Path History Register (PHR) is used to record branch history, the pipeline breaks down branch history recovery into three steps: first, it measures the PHR length; then, it identifies the range of input bits of the PHR; and lastly, it enumerates the location of each input bit. Subsequently, it narrows the input range of the PHT by guessing the PHT associativity using low PC bits and probing PHR bits in the index and tag functions. However, the existing pipeline has the following major issues:

\begin{figure}[htbp]
  \centering
  \includegraphics[width=0.9\linewidth]{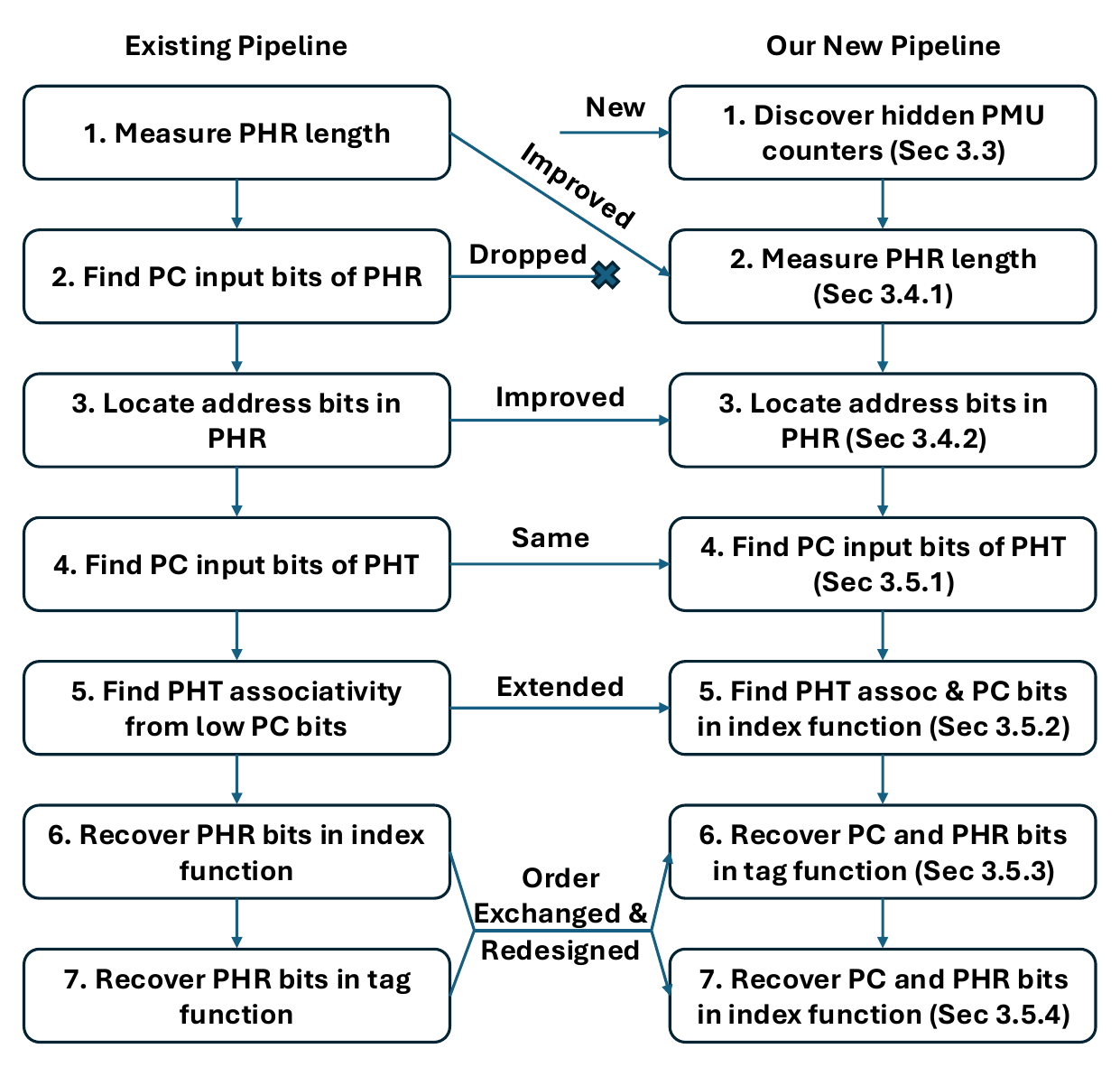}
  \caption{Reverse engineering pipeline: our improvements over the existing pipeline}
  \label{fig:pipeline}
\end{figure}

\begin{itemize}
  \item \textbf{Interference of mispredictions between different branches}: Existing methods primarily use conditional branches to set the PHR to known values. However, these branches also contribute to branch mispredictions and introduce noise to the measured performance monitoring counters (PMC). This makes it hard for these methods to deal with complex CBP designs. To overcome this, we adopt a novel approach of utilizing indirect branches to inject values into the PHR and then record conditional branch mispredictions using PMC. Since PMC recording conditional branch mispredictions is not always publicly available (e.g., in some AMD, ARM, and Qualcomm microarchitectures), we add an additional step to recover hidden counters.
  \item \textbf{Clearing and setting PHR}: Existing methods rely on clearing the PHR to zero and setting it to arbitrary values. However, this is challenging when the PHR uses many PC bits, as in Firestorm and Oryon. We rewrite existing microbenchmarks to inject PHR in a differential way, where two PHRs differ by only one bit. This eliminates the need to find PC inputs to clear the PHR.
  \item \textbf{Simple assumptions about PHT index and tag functions}: The existing pipeline makes assumptions about the simpler PHT index and tag functions of Intel, which do not hold for Firestorm and Oryon. For instance, Intel uses only one PC bit in the index function, whereas Firestorm and Oryon use multiple. Therefore, we propose a comprehensive approach to find the associativity and all PC bits in the index function together.
  \item \textbf{The order of index and tag function recovery}: The tag function should be recovered before the index function because creating set conflicts requires prior knowledge of the tag function. The existing pipeline uses assumptions about the index function organization, which only applies to Intel CBP. Instead of making assumptions, we recover them in the correct order.
\end{itemize}

To address these deficiencies, we propose a new pipeline in the right part of Figure \ref{fig:pipeline} to successfully recover the CBPs of Firestorm and Oryon. This pipeline is more general with fewer assumptions, making it applicable to ARM and Intel CBPs as well.

\subsection{Experimental Setup}

To reverse engineer the conditional branch predictors, we utilize the Apple M1 processor for the Apple Firestorm microarchitecture and the Qualcomm X1E-80-100 processor for the Qualcomm Oryon microarchitecture. We conduct our experiments in bare-metal Linux environments (e.g. Asahi Linux on Apple M1) to access performance monitoring counters directly.

\subsection{Discover Hidden PMCs}
\label{sec:hidden_pmu}

As previously mentioned, this is one of the prerequisites for the subsequent reverse engineering of the more complex CBPs.
There are three involved counters that record different types of branch mispredictions: the all branch counter, the conditional branch counter, and the indirect branch counter.

According to open literature, Apple provides all the three counters, and the numbers are respectively 0xcb, 0xc5, and 0xc6. However, Qualcomm only publishes the first one, 0x10 (or 0x22) \cite{armv8}.

Therefore, we design three microbenchmarks that have a known number of mispredictions for each type of branch. We then enumerate the hidden performance counters to determine if the actual values match our expectations. The first microbenchmark contains conditional branches whose directions are based on random numbers, resulting in a conditional branch misprediction rate of 50\%. The second microbenchmark contains indirect branches that jump to two target addresses randomly, leading to an indirect branch misprediction rate of 50\%. The third microbenchmark combines the first two microbenchmarks by running both conditional and indirect branches simultaneously.

When executing the loop body $n$ times, we anticipate that three counters recording different types of branch mispredictions will yield the values presented in Table \ref{tbl:pmu}. We have enumerated the PMC space and identified the corresponding counters as listed in Table \ref{tbl:pmu_found}. Our findings are: for Qualcomm Oryon, the counters 0x10 and 0x22 correspond to the standard counters defined in ARMv8 \cite{armv8}, but the counters 0x400, 0x80d, and 0x80e remain undisclosed. 
In the subsequent steps, we utilize the 0xc5 counter for Apple Firestorm and the 0x400 counter for Qualcomm Oryon to measure conditional branch misprediction rates. This approach allowed us to inject branch histories via indirect branches without introducing noise to the performance counters, which is required for subsequent microbenchmarks.

\begin{table}[htbp]
  \centering
  \caption{Expected branch misprediction counter values running the three microbenchmarks (abbreviated as MB below)}
  \label{tbl:pmu}
  \begin{tabular}{cccc}
      \toprule
      \textbf{Counter} & \textbf{MB1} & \textbf{MB2} & \textbf{MB3} \\
      \midrule
      All & $n/2$ & $n/2$ & $n$ \\
      Conditional & $n/2$ & $0$ & $n/2$ \\
      Indirect & $0$ & $n/2$ & $n/2$ \\
      \bottomrule
  \end{tabular}
\end{table}

\begin{table}[htbp]
  \centering
  \caption{Reverse engineered branch misprediction counters in Apple Firestorm and Qualcomm Oryon (\underline{underline}: previously undisclosed)}
  \label{tbl:pmu_found}
  \begin{tabular}{ccc}
      \toprule
      \textbf{Counter} & \textbf{Apple Firestorm} & \textbf{Qualcomm Oryon} \\
      \midrule
      All & 0xcb & 0x10/0x22 \\
      Conditional & 0xc5 & \underline{0x400} \\
      Indirect & 0xc6 & \underline{0x80d/0x80e} \\
      \bottomrule
  \end{tabular}
\end{table}

\subsection{Reverse Engineering The Path History Register}

The PHR records recent taken branches by shifting a certain number of bits and then XOR-ing the footprint generated by the branch and the target address of the taken branch. To reverse engineer the PHR, we need to recover the length of the register, the shift amount upon each taken branch, and the method by which the footprint is generated.

\subsubsection{Measure PHR Length}

First, we measure the number of taken branches that can be tracked in the PHR using an approach improved from that described in \cite{halfhalf} with the newly discovered hidden PMC. We inject a variable number of unconditional branches, referred to as dummy branches, to shift the PHR. We create two random but correlated branches with various dummy branches in between. If the CBP sees the history of the first correlated branch when predicting the direction of the second one, it can predict correctly; otherwise, it has a 50\% chance of mispredicting. Since we have specific PMCs for conditional branches alone, we use an indirect branch for the first correlated branch and a conditional branch for the second one. The pseudo-code is shown in Listing \ref{lst:phr_length}, which is similar to Listing 3 of \cite{halfhalf}, but with the first conditional branch replaced by an indirect branch.

\begin{figure}[htbp]
  \begin{lstlisting}[label=lst:phr_length,caption=Microbenchmark to measure how many taken branches an be tracked in the PHR]
for (int i = 0;i < n;i++) {
  int k = rand() % 2; // random direction
  // indirect branch with randomized target
  goto (k ? l1 : l0);
  l0: asm ("nop"); l1:
  // repeated unconditional jumps (dummy branches)
  goto l2; l2: goto l3; l3: // ... omitted
  if (k) {} // conditional branch
}
  \end{lstlisting}
\end{figure}

\begin{figure}[htbp]
  \centering
  \includegraphics[width=0.9\linewidth]{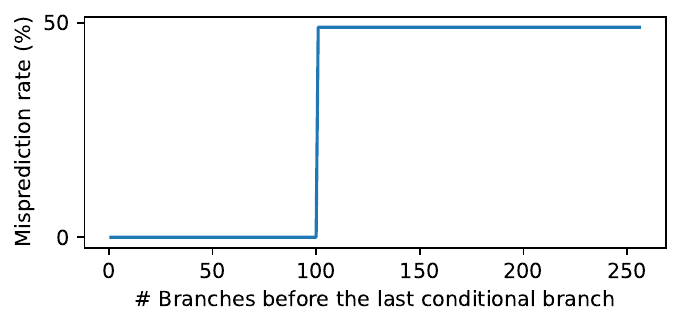}
  \caption{Misprediction rate of the last conditional branch due to different number of branches before it}
  \label{fig:phr_size}
\end{figure}

We run the microbenchmark on both microarchitectures and obtain identical results, as depicted in Figure \ref{fig:phr_size}. The misprediction rate increases from 0\% at 100 branches to 50\% at 101 branches. This indicates that the PHR in both microarchitectures can only record information from the most recent 100 taken branches. The result also confirms that the global history register (GHR) is not used in place of the PHR: if the GHR were used solely for prediction, it would only contain the direction history of the conditional branch, preventing the conditional branch predictor from achieving a 0\% misprediction rate.

\subsubsection{Locate address bits in PHR}

Next, we reverse engineer the update mechanism and bit width of the PHR. We begin by determining the location of the branch address within the PHR. Given that the width of the branch footprint may exceed the shift amount, some footprint bits may be shifted out of the PHR earlier than others. To investigate this, we inject the branch direction into different bits of the branch address to observe how many branches are required to shift these bits out of the PHR. 

Unlike the approach in \cite{halfhalf}, which involves finding a method to clear the PHR, we employ a differential approach. This involves using two PHRs that differ by only one bit while the other bits can be non-zero. We use a pair of branches with a single bit difference in their branch addresses but identical target addresses to inject a single bit, as illustrated in Listing \ref{lst:phrb_loc}. In this scenario, the first conditional branch will contribute approximately 25\% to the overall conditional branch misprediction rate.

\begin{figure}[htbp]
  \begin{lstlisting}[label=lst:phrb_loc,caption=Microbenchmark to find the locations of branch addresses in footprint]
for (int i = 0;i < n;i++) {
  int k = rand() % 2; // random direction
  // two branches with same target address and one bit difference in branch address
  if (k) goto l0; // conditional branch
  goto l0; // unconditional branch
  l0:
  // repeated unconditional jumps omitted
  if (k) {} // conditional branch
}
  \end{lstlisting}
\end{figure}

\begin{figure}[htbp]
  \centering
  \begin{subfigure}[b]{\columnwidth}
      \centering
      \includegraphics[width=\linewidth]{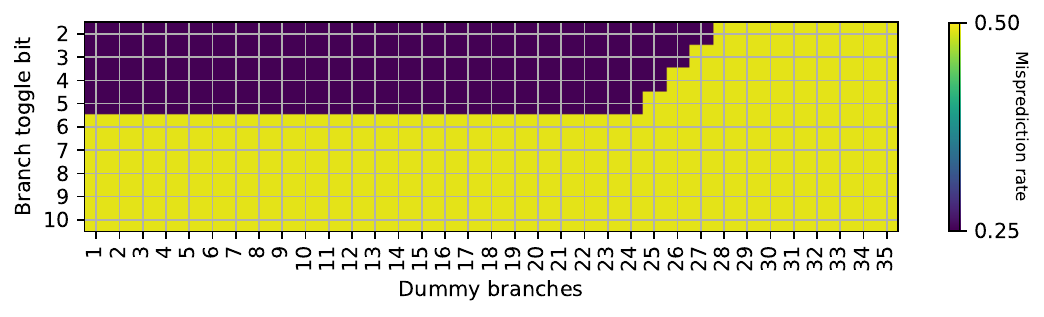}
      \caption{Apple Firestorm}
  \end{subfigure}
  \begin{subfigure}[b]{\columnwidth}
      \centering
      \includegraphics[width=\linewidth]{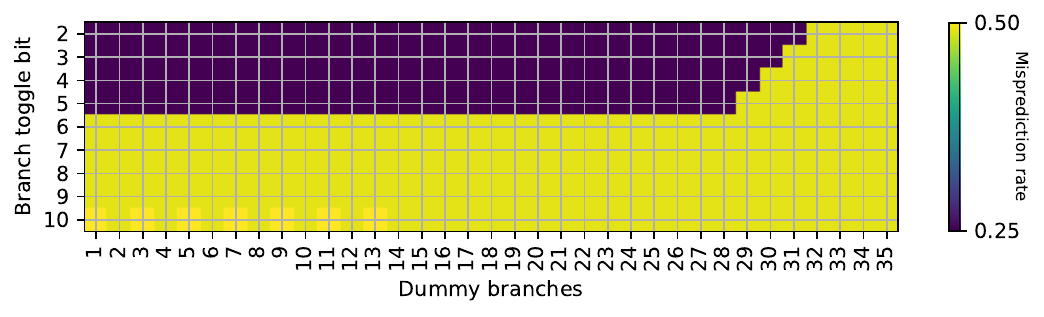}
      \caption{Qualcomm Oryon}
  \end{subfigure}
  \caption{Misprediction rate due to different number of dummy branches and branch address toggle bit}
  \label{fig:phrb_loc}
\end{figure}

We run the microbenchmark on both microarchitectures, and the results are presented in Figure \ref{fig:phrb_loc}. The findings indicate that, for both Firestorm and Oryon, only the bits 5:2 of the branch address (abbreviated as B below) are utilized in the computation of the footprint. The remaining bits of the branch address do not contribute to the PHR footprint computation. Each bit corresponds to a single number of dummy branches, which strongly suggests that the shift amount of the PHR is one.

Subsequently, we employ a similar differential approach to identify the target address bits, utilizing the newly discovered hidden PMC. Instead of using a branch pair to inject the branch address, we utilize an indirect branch with variable target addresses. To bypass the numerous NOP instructions between the two target addresses, we copy the first dummy branch to the first target address to jump over the NOP instructions. We maintain the same B[5:2] for both branches to avoid altering the PHR. To circumvent the limitations of assemblers and linkers, we utilize just-in-time (JIT) compilation to place instructions directly into memory.

\begin{figure}[htbp]
  \centering
  \includegraphics[width=\linewidth]{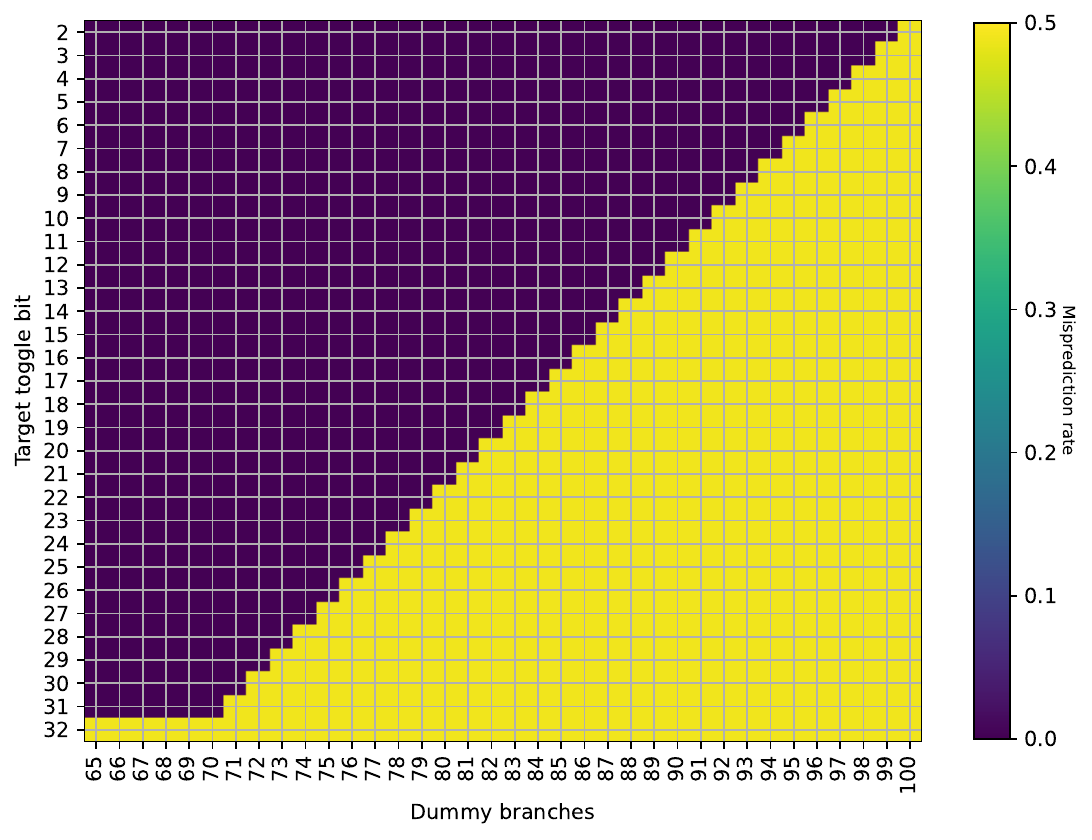}
  \caption{Misprediction rate due to different number of dummy branches and target address toggle bit}
  \label{fig:phrt_loc}
\end{figure}

The results are illustrated in Figure \ref{fig:phrt_loc}. Both Firestorm and Oryon exhibit identical behavior in the target address injection test: the bits 31:2 of the target address (abbreviated as T below) are utilized in the PHR footprint computation. Similar to the branch address, the target address is shifted into the PHR by one bit for each taken branch. Therefore, we conclude that the shift amount of the PHR is one bit. Given that the PHR can track 100 taken branches, we infer that the PHR has a length of 100 bits.

Finally, we recover the footprint computation function of the PHR. Given that we already know the PHR shift amount and bit width, we can convert the maximum number of dummy branches that allow perfect prediction of each B/T bits to their respective locations within the footprint. Based on Figure \ref{fig:phrb_loc} and Figure \ref{fig:phrt_loc}, if both the branch and target addresses are combined into a single PHR register, it will result in a large gap within the footprint. Therefore, it is plausible that the PHR is split into two registers: one called PHRT (Path History Register for Target address), which exclusively records the target address, and another called PHRB (Path History Register for Branch address). The update functions for both PHRT and PHRB are as follows for Firestorm and Oryon:

\begin{align*}
  \mathrm{PHRT}_{\mathrm{new}} &= (\mathrm{PHRT}_{\mathrm{old}} \ll 1) \oplus \mathrm{T}[31:2] \\
  \mathrm{PHRB}_{\mathrm{new}} &= (\mathrm{PHRB}_{\mathrm{old}} \ll 1) \oplus \mathrm{B}[5:2] \\
\end{align*}

Subsequently, Firestorm and Oryon differ only in the width of PHRB: Oryon features a 32-bit PHRB, whereas Firestorm has a 28-bit PHRB. The PHRT remains consistent at 100 bits for both microarchitectures. In the following section, we will demonstrate that the split PHRT and PHRB approach is indeed employed in the actual hardware.

\subsection{Recover the 1st PHT}

In this section, we reverse-engineer the PHTs used in the TAGE algorithm. Currently, we do not know the number of PHTs and the history lengths of each table, so we reverse-engineer the PHT with the longest history (referred to as the 1st PHT below) first. To force the 1st PHT to be used, we inject a random direction $k$ into PHR[99], which is the highest bit of the PHR, and correlate it with the conditional branches in microbenchmarks. Since Firestorm and Oryon differ in PHT details, but the same reverse-engineering techniques can be applied, we primarily present Firestorm results in the following paragraphs and we will publish the full results online.

\subsubsection{Find PC inputs of PHT}

First, we need to determine the inputs to the PHT. As previously mentioned, the program counter (PC) of the conditional branch and the PHR are input values to the PHT, but we do not yet know which bits of the PC are utilized. Therefore, we design a microbenchmark to observe whether the conditional branch predictor can distinguish between two conditional branches whose addresses differ by only one bit. Instead of clearing the PHR as in \cite{halfhalf}, we use a long chain of dummy branches to reset the PHR to an unknown but constant value. Subsequently, two conditional branches are predicted by the CBP using the same PHR. If the bit difference between the two conditional branches does not affect the PHT index or tag computation, the CBP will treat them as a single branch and fail to predict correctly. We run the microbenchmark on both microarchitectures and conclude that the Firestorm CBP uses PC[18:2] as input, while the Oryon CBP uses PC[12:2].

\subsubsection{Find PHT associativity and PC bits in index function}

Next, we need to determine whether the PC bits are used in the index or tag function of the PHTs. In Intel Alder Lake, only one PC bit was used in the index, while the others were used in the tag \cite{halfhalf}. This is no longer the case in Firestorm and Oryon CBP. To identify the PC bits that belong to the index function, we use one PHR to predict several branches with different PC values. We place the branches at multiples of a power-of-two base address. These addresses may differ in several PC bits. If all these bits only reside in the tag function, then the branches will belong to one set of PHT, and the maximum number of branches that can be predicted accurately will equal the associativity. If some bits appear in the index function, then the branches will be scattered across multiple sets, and we can observe a power-of-two factor multiplied to the maximum number of branches predicted accurately. For example, if only PC[6] and PC[9] are included in the index function, while others are in the tag function, when we use different base addresses, we should be able to observe the following results, assuming a 4-way set associative PHT on Firestorm:

\begin{itemize}
  \item \textbf{Case 1. All PC bits reside in tag}: Using a base address of $2^3$ bytes, the first five branches are placed at 0x8, 0x10, 0x18, 0x20, 0x28 addresses respectively. They are mapped to the same index, so adding the fifth branch will lead to branch misprediction. Thus, the maximum number of branches without mispredictions would be four. For base address of $2^{10}$ to $2^{17}$ bytes, the result is the same.
  \item \textbf{Case 2. Only one PC bit reside in index}: Using base address of $2^4$ bytes, the first four branches are mapped to the same index, but the fifth branch goes to a new set. So another set of four branches can be predicted without mispredictions. The ninth branch will lead to set conflict. Using base address of $2^5$ or $2^7$ to $2^9$ bytes, the result is the same.
  \item \textbf{Case 3. Two PC bits reside in index}: Using base address of $2^6$ bytes, now the branches can span both PC[6] and PC[9] index bits to achieve sixteen branches without mispredictions using four sets of branches.
  \item \textbf{Case 4. PC bits go beyond the input range}: Using base address of $2^{18}$ and $2^{19}$, the result is two or one branches because higher bits are no longer computed in index or tag function.
\end{itemize}

We design a microbenchmark to observe the actual numbers, utilizing the newly discovered hidden PMC. Then, we need to let the CBP use the same PHR to predict multiple branches. If we use an indirect branch to jump to these conditional branches, the indirect branch itself will introduce PHR differences. Therefore, we use two indirect branches to maintain the same PHR while landing at different conditional branches. For example, if we want to jump to a $2^n$ address, assuming $n-1>5$:

\begin{itemize}
  \item \textbf{Branch \#1}: One indirect branch to jump to a $2^{n-1}$ address: B=0, T[n-1]=1, then $PHR_1=(PHR_0 \ll 1) \oplus (1 \ll (n-3))$
  \item \textbf{Branch \#2}: For every $n$, one indirect branch at $2^{n-1}$ to jump to $2^n$: B[n-1]=1, T[n]=1, since $n-1>5$, then $PHR_2=(PHR_1 \ll 1) \oplus (1 \ll (n-2)) = PHR_0 \ll 2$
\end{itemize}

\begin{figure}[htbp]
  \centering
  \includegraphics[width=\linewidth]{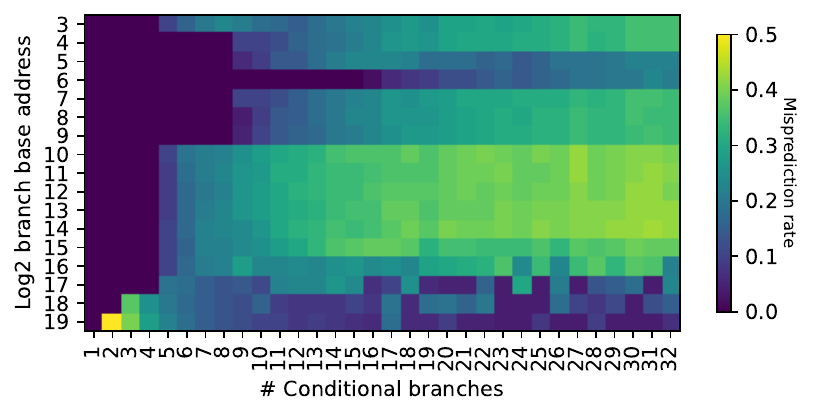}
  \caption{Misprediction rate due to different number of conditional branches and log2 of branch base address}
  \label{fig:pht_assoc}
\end{figure}

The PHR changes of the two indirect branches via target branch cancel out. If $n-1 \le 5$, we fill the space between the targets of the first indirect branch with NOPs, and then use a second indirect branch to cancel out the contribution of T bits of the first indirect branch. Either way, we can let the CBP predict different branches using the same PHR. This technique can be extended to use three or more branches to inject PHR via branch address in other microbenchmarks. The results of the microbenchmark on Firestorm are shown in Figure \ref{fig:pht_assoc}.

The results matched our expected outcomes based on the assumption that only PC[6] and PC[9] are used in the index function and the PHT is 4-way set associative. Therefore, we can conclude that the 1st PHT of the Firestorm CBP is 4-way set associative and uses PC[6] and PC[9] in its index function. Similarly, we find that the 1st PHT of the Oryon CBP is also 4-way associative but uses PC[6] and PC[7] in its index function instead.

\subsubsection{Recover PC and PHR bits in tag function}

Next, we recover the tag function. To identify which bits are XOR-ed in the tag function, we use two independent random variables: $k$ and $l$. We inject $k$ and $l$ into different bits of the PHR or PC and let the CBP predict a conditional branch conditioned by $k \oplus l$. If the two bits are XOR-ed in the tag function and do not appear in the index function, misprediction will occur. We enumerate all pairs among the PHR bits (using the target address and branch address separately) and PC bits. On Firestorm, 12 groups of XOR-ed bits can be observed (PHRT denotes injecting PHR via the target address, PHRB for injecting with the branch address, PHRT[a,b,...,d] means PHRT[a] $\oplus$ PHRT[b] $\oplus \cdots \oplus$ PHRT[d], PHRB likewise):

\begin{enumerate}
  \item PC[7] $\oplus$ PHRT[0,12,...,96] $\oplus$ PHRB[8,21]
  \item PC[8] $\oplus$ PHRT[1,13,...,97] $\oplus$ PHRB[9,22]
  \item PC[9] $\oplus$ PHRT[2,14,...,98] $\oplus$ PHRB[10,23]
  \item PC[10] $\oplus$ PHRT[3,15,...,87] $\oplus$ PHRB[11,12,24,25]
  \item PC[11] $\oplus$ PHRT[4,16,...,88] $\oplus$ PHRB[0,13,26]
  \item PC[12] $\oplus$ PHRT[5,17,...,89] $\oplus$ PHRB[1,14,27]
  \item PC[13] $\oplus$ PHRT[6,18,...,90] $\oplus$ PHRB[2,15]
  \item PC[14] $\oplus$ PHRT[7,19,...,91] $\oplus$ PHRB[3,16]
  \item PC[15] $\oplus$ PHRT[8,20,...,92] $\oplus$ PHRB[4,17]
  \item PC[16] $\oplus$ PHRT[9,21,...,93] $\oplus$ PHRB[5,18]
  \item PC[17] $\oplus$ PHRT[10,22,...,94] $\oplus$ PHRB[6,19]
  \item PC[18] $\oplus$ PHRT[11,23,...,95] $\oplus$ PHRB[7,20]
\end{enumerate}

The results are consistent with our previous observations. Since PHRB and PHRT participate in the tag function in different ways, we can deduce that the PHR is split into two registers: PHRT and PHRB. Because PC[5:2] are also included in the index or tag function and do not reside in the index function, they appear in the tag function without being XOR-ed. We conclude that the tag function of the 1st PHT contains 16 bits: 12 bits from the XOR group above and 4 bits from PC[5:2].

\subsubsection{Recover PC and PHR bits in index function}

Lastly, we recover the index function of the 1st PHT. Since we already know the tag function, we can create multiple branches with different PC[5:2] so that they will never collide in the tag. We also know the associativity, so we use two groups of conditional branches with 4 branches each. Again, we inject two random variables $k$ and $l$ into bits of PHRB, PHRT or PC, and let the conditional branches be conditioned by $k \oplus l$. Consider the different cases where the bits reside:

\begin{itemize}
  \item If $k$ and $l$ are both injected into bits in the index function but not XOR-ed, the CBP can predict all eight branches without misprediction since they are mapped to different sets.
  \item If $k$ and $l$ are both injected into bits in the index function and XOR-ed, the CBP can predict at most four branches without misprediction because they are mapped to the same set.
  \item Otherwise, the CBP can predict at most two branches without misprediction because both directions of the same branch ($k \oplus l = 0$ and $k \oplus l = 1$) will occupy an entry in one set.
\end{itemize}

Thus, we are able to identify all index bits and their XOR relationships. On Firestorm, 10 index bits are discovered for the 1st PHT:

\begin{enumerate}
  \item PHRT[2] $\oplus$ PHRT[43] $\oplus$ PHRT[93]
  \item PHRT[7] $\oplus$ PHRT[48] $\oplus$ PHRT[99]
  \item PHRT[12] $\oplus$ PHRT[63] $\oplus$ PHRB[5]
  \item PHRT[17] $\oplus$ PHRT[68] $\oplus$ PHRB[10]
  \item PHRT[22] $\oplus$ PHRT[73] $\oplus$ PHRB[15]
  \item PHRT[27] $\oplus$ PHRT[78] $\oplus$ PHRB[20]
  \item PHRT[33] $\oplus$ PHRT[83] $\oplus$ PHRB[25]
  \item PHRT[38] $\oplus$ PHRT[88] $\oplus$ PC[9]
  \item PHRT[53] $\oplus$ PHRT[58] $\oplus$ PHRB[0]
  \item PC[6]
\end{enumerate}

\subsubsection{Conclusion}

We have recovered the index and tag functions of the 1st PHT of Firestorm, which is 4-way 4096-entry in capacity. We have also demonstrated that the Firestorm CBP uses two separate registers, PHRB and PHRT, to record branch history.

\subsection{Recover the rest PHTs}

After recovering the 1st PHT, which has the longest history length, we need to recover the remaining PHTs with shorter history lengths. Due to the nature of how TAGE works, upon mispredictions in a PHT with a shorter history length, new entries will be allocated in a PHT with a longer history length \cite{tage}. Therefore, we need to remove the effect of PHTs with longer history lengths to recover the rest of the PHTs. In \cite{halfhalf}, to recover the base predictor, they zeroed out the PHR so that the index is cleared to zero, and then pre-filled the PHT with branches. This technique can also be applied to recovering PHTs: for example, if we inject $k$ into PHRT[43] in the 2nd PHT (the PHT with the second longest history length), we also inject $k$ into PHRT[93] so that they get XOR-ed in the index function of the 1st PHT. This method works well for the first few PHTs but has some limitations: clearing out index bits may not always be applicable. For example, if an index bit only uses very low PHR bits, such as PHRT[0] $\oplus$ PHRT[1] $\oplus$ PHRB[2], then PHTs with shorter history lengths will be involved. This was actually observed while recovering the 3rd PHT of Firestorm. To make matters worse, zeroing out an index for one bit introduces another, while the new bit may require a third bit to clear out in another PHT, essentially creating a chain reaction.

To overcome these limitations, we introduce a new approach that inserts independent random variables into history bits that do not belong to the current PHT. For example, to recover the 2nd PHT of Firestorm, which uses 57 bits of PHRT, we inject random values into PHRT[99:57]. Even if new entries are allocated in the 1st PHT, there is little possibility of having a tag hit later. If without the hidden PMC, this approach will no longer work due to interference. Combining these techniques, we recover the remaining PHTs of Firestorm and Oryon. From Table \ref{tbl:phts}, the size of the Oryon CBP, considering tag bits, is computed as $(4 \times 2^{10} \times 3 + 4 \times 2^{11} \times 2 + 6 \times 2^{11}) \times 16 = 655360$ bits, which equals the 80KB capacity announced by Qualcomm in \cite{hotchips}.

\begin{table}[htbp]
  \centering
  \caption{History length, associativity and the number of index bits of PHTs (*: low confidence)}
  \label{tbl:phts}
  \begin{tabular}{ccccc}
      \toprule
      \textbf{uArch} & \textbf{PHRT} & \textbf{PHRB} & \textbf{Assoc} & \textbf{\# Index bits} \\
      \midrule
      \multirow{6}{4em}{Firestorm} & 100 & 28 & 4 & 10 \\
      & 57 & 28 & 4 & 10 \\
      & 32 & 28 & 4 & 10 \\
      & 18 & 18 & 4 & 11 \\
      & 11 & 11 & 6 & 11 \\
      & 6 & 6 & 6 & 11* \\
      \midrule
      \multirow{6}{4em}{Oryon} & 100 & 32 & 4 & 10 \\
      & 52 & 32 & 4 & 10 \\
      & 27 & 27 & 4 & 10 \\
      & 14 & 14 & 4 & 11 \\
      & 7 & 7 & 4 & 11* \\
      & 4 & 4 & 6 & 11* \\
      \bottomrule
  \end{tabular}
\end{table}

\section{Optimization with CBP Model}

In this section, we observe two previously undisclosed effects that may impair branch prediction accuracy: \textbf{Scatter} and \textbf{Annihilation}. Utilizing the findings from reverse engineering, we develop models of the CBPs for Firestorm, Oryon and Intel. These models enable us to explain the underlying mechanisms of the two effects and subsequently enhance performance.

\subsection{Scatter Effect}

\subsubsection{Phenomenon}

We observe a scenario where the insertion of a single NOP instruction can simultaneously reduce both the branch misprediction rate and the Mispredictions Per Kilo Instructions (MPKI), thereby enhancing performance. This scenario can be demonstrated through an example, that is, the execution of a typical binary search program using keys that adhere to a Zipf distribution with an exponent of, say, 0.9. The corresponding code and generated assembly are presented in Listings \ref{lst:binary_search_code} and \ref{lst:binary_search_asm}, respectively.

\begin{figure}[htbp]
  \begin{lstlisting}[label=lst:binary_search_code,caption=A typical binary search program]
int binary_search(int *table, int size, int key) {
  int low = 0, high = size - 1;
  while (low <= high) {
    int probe = (low + high) / 2;
    int v = table[probe];
    if (v < key) {
      low = probe + 1;
    } else if (v > key) {
      high = probe - 1;
    } else {
      return probe;
    }
  }
  return -1;
}
  \end{lstlisting}
\end{figure}

\begin{figure}[htbp]
  \begin{lstlisting}[label=lst:binary_search_asm,style=customasm,caption=Compiled assembly of the binary search program]
binary_search: % start address is aligned
      subs    w9, w1, #1
      b.lt    .L6
      mov     w10, wzr
      b       .L3
.L2:
      add     w10, w8, #1
      cmp     w10, w9
      b.gt    .L6
.L3:
      add     w8, w10, w9
      lsr     w8, w8, #1
      ldr     w11, [x0, w8, uxtw #2]
      cmp     w11, w2
      b.lt    .L2
      b.le    .L7
      sub     w9, w8, #1
      cmp     w10, w9
      b.le    .L3
.L6:
      mov     w8, #-1
.L7:
      mov     w0, w8
      ret
  \end{lstlisting}
\end{figure}

\begin{figure}[htbp]
  \centering
  \includegraphics[width=\linewidth]{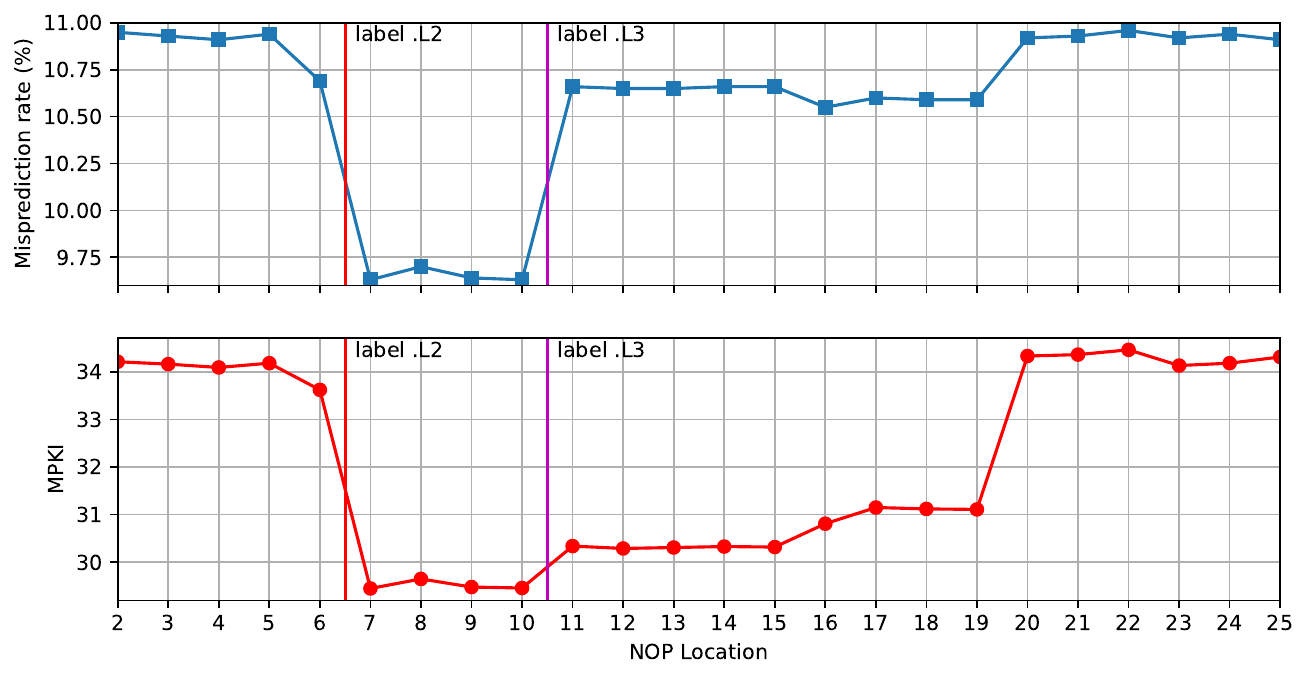}
  \caption{Branch misprediction rate and MPKI due to different NOP insert location on Oryon}
  \label{fig:conflict}
\end{figure}

By inserting a single NOP instruction into the assembly, we observe a significant reduction in the branch misprediction rate, dropping from 10.92\% to 9.63\%, and a corresponding decrease in MPKI from 34.5 to 29.5 on the Oryon microarchitecture. This optimization finally results in a 7\% speedup on Oryon. The placement of the NOP instruction is crucial, as illustrated in Figure \ref{fig:conflict}. The minimum misprediction rate and MPKI are attained when the NOP instruction is inserted between the labels \verb|.L2| and \verb|.L3|.

Moreover, this effect is universal, that is, it also exists in Intel and Apple processors, and the same optimization results in a 6\% speedup on Intel Cascade Lake and a 2\% speedup on Firestorm. 

\begin{table}[htbp]
  \centering
  \caption{Branch and target address of the two critical branches (b.lt .L2 and b.le .L3) due to different NOP locations}
  \label{tbl:branch_addr}
  \begin{tabular}{ccccc}
      \toprule
      \textbf{NOP location} & \textbf{b.lt B} & \textbf{b.lt T} & \textbf{b.le B} & \textbf{b.le T} \\
      \midrule
      Not inserted & 0x2c & 0x10 & 0x3c & 0x1c \\
      from \verb|.L2| to \verb|.L3| & 0x30 & 0x10 & 0x40 & 0x20 \\
      from \verb|.L3| to \verb|b.lt .L2| & 0x30 & 0x10 & 0x40 & 0x1c \\
      \bottomrule
  \end{tabular}
\end{table}

\subsubsection{Analysis and conclusion}

With the branch predictor model at our disposal, we can now explain the underlying reasons for these observations. The performance-critical conditional branches within the loop are \verb|b.lt .L2| and \verb|b.le .L3|. These branches are data-dependent and correlated due to the Zipf distribution. According to the model, the tag function of the Oryon CBP XORs the PHRB and PHRT bits with zero offset. When a NOP instruction is inserted between the labels \verb|.L2| and \verb|.L3|, as shown in Table \ref{tbl:branch_addr}, the contributions of the two branches to the PHT tag differ by only one bit. In contrast, other NOP locations or the absence of a NOP result in more significant differences in the PHT tag. Additionally, the two conditional branches differ in PC[5:2], thus avoiding any aliasing. Given the different TAGE entries for the same branch due to varying PHRs, fewer entries are occupied because the tag function exhibits less randomness, allowing TAGE to be trained more rapidly. Without the additional NOP or if the NOP is not properly placed, one branch may be mapped to more TAGE entries, leading to increased capacity conflicts.

In conclusion, the \textbf{Scatter} effect refers to the phenomenon where a single conditional branch is distributed across a large number of TAGE PHT entries, resulting in higher MPKI and degraded performance. Our analysis indicates that this effect can be mitigated by examining the PHT tag bits contributed by the critical branches and strategically inserting NOPs.

\subsection{Annihilation Effect}

\subsubsection{Phenomenon}

Based on the CBP model, we observe that numerous taken branches share identical footprints, implying that the CBP can no longer differentiate between these branches, thereby leading to increased mispredictions. We term this phenomenon the \textbf{Annihilation} effect. To delve deeper into this issue, we examine the \verb|leela_r| benchmark from SPEC INT 2017 using distinct PHR update functions from Intel, Apple and Qualcomm CBPs. We conduct a static analysis of all branch pairs within the same function, and count the number of branch pairs whose PHR footprints collide for each CBP. The results are presented in Table \ref{tbl:annihilation}, indicating that the Alder Lake CBP generates a smaller number of collisions than Firestorm, Oryon and Skylake whereas Haswell performs the worst.

\begin{table}[htbp]
  \centering
  \caption{Comparison of different CBPs on Annihilation effect}
  \label{tbl:annihilation}
  \begin{tabular}{ccc}
      \toprule
      \textbf{uArch} & \textbf{PHR Input} & \textbf{\# Collisions}  \\
      \midrule
      Haswell & B[19:4] and T[5:0] & 270 \\
      Firestorm/Oryon & B[5:2] and T[31:2] & 138 \\
      Skylake & B[18:3] and T[5:0] & 134 \\
      Alder Lake & B[15:0] and T[5:0] & 44 \\
      \bottomrule
  \end{tabular}
\end{table}

\subsubsection{Analysis and conclusion}

Regarding the PHR update function, Intel and ARM adopt the more traditional approach of using a single PHR for both branch and target addresses \cite{halfhalf, trustzonetunnel, phr}. In contrast, Apple and Qualcomm employ two separate registers with a predominant width of footprint in T[31:2]. This approach may present challenges in speculative PHR update and restoration but can mitigate the occurrence of the \textbf{Annihilation} effect with regard to target address. However, Apple and Qualcomm only use a small portion of the branch address bits in the PHR footprint. Therefore it becomes relatively easy for collisions to occur in typical programming patterns such as \verb|goto end| for cleanups and \verb|continue| statements within loop bodies, as in these patterns, branches often jump to the same target address. Conversely, if lower branch address bits are not used, as in Intel Haswell and Skylake, it is easy for two adjacent branches to generate the same footprint. Intel improved this in Alder Lake by utilizing lower branch address bits. 

In conclusion, the \textbf{Annihilation} effect of colliding PHR footprints can be mitigated through redesigning the PHR update function to include lower bits of both the branch and target addresses.

\section{Comparing CBPs of Different CPU Vendors}

In this section, we conduct a comparative analysis of the conditional branch predictors (CBPs) from various CPU vendors, employing both quantitative and qualitative approaches. We implement these CBPs on a unified branch predictor model and evaluate their branch prediction accuracy using SPEC INT 2017 and Geekbench 5 benchmarks. Subsequently, we analyze the design considerations underlying these implementations.

\subsection{Branch Prediction Accuracy of CBPs}
\label{sec:mpki}

\subsubsection{Comparing the actual CBPs}

\begin{figure}[htbp]
  \centering
  \includegraphics[width=\linewidth]{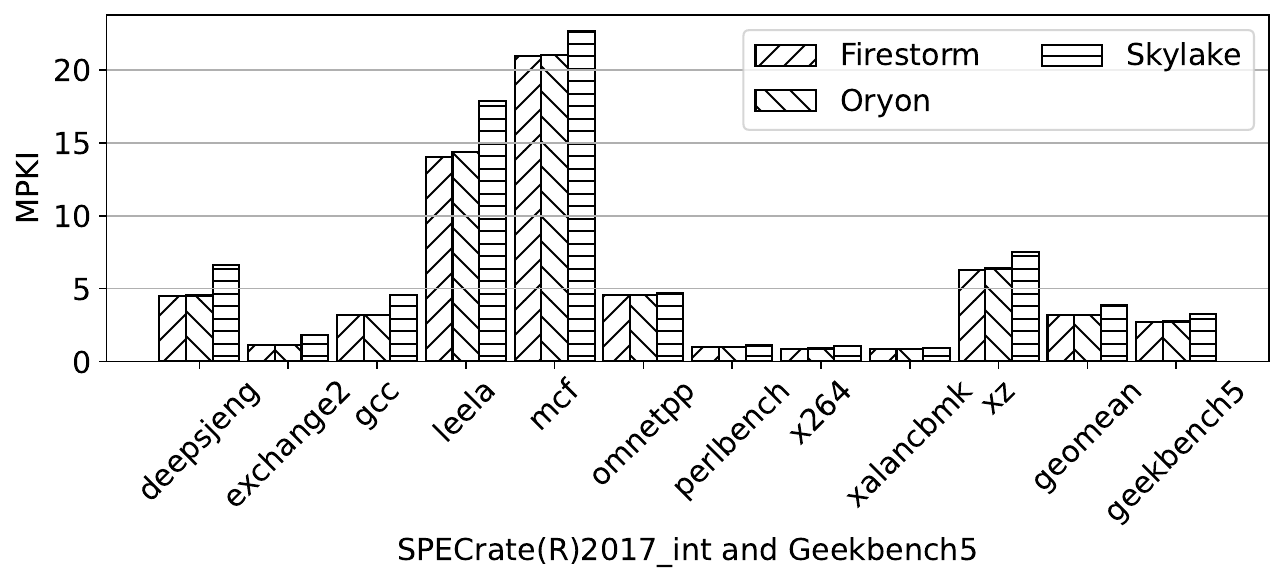}
  \caption{MPKI of each CBP on SPEC INT 2017 and Geekbench 5 benchmarks}
  \label{fig:spec}
\end{figure}

Given that the CBPs of Intel \cite{halfhalf}, Apple, and Qualcomm cores have been recovered (as shown in Table \ref{tbl:cbps}), we implement them on an in-house ChampSim-like \cite{champsim} standalone branch predictor model that aligns with a commercial processor. We simulate these CBPs using SPEC INT 2017 and Geekbench 5 benchmarks compiled for ARM64, employing the SimPoint \cite{simpoint} methodology. The results, depicted in Figure \ref{fig:spec}, indicate that the Firestorm CBP performs the best, narrowly outperforming the Oryon CBP by 1\%, while the Skylake CBP significantly lags behind by more than 20\%.

\begin{table}[htbp]
  \centering
  \caption{TAGE configuration of CBPs}
  \label{tbl:cbps}
  \begin{tabular}{cccc}
      \toprule
      \textbf{uArch} & \textbf{PHR Width} & \textbf{\# PHT} & \textbf{\# Entries}  \\
      \midrule
      Apple Firestorm & $100+28$ & 6 & 44K \\
      Qualcomm Oryon & $100+32$ & 6 & 40K \\
      Intel Skylake & $186$ & 3 & 6K \\
      \bottomrule
  \end{tabular}
\end{table}

\subsubsection{Comparing the CBPs under the same capacity}

Given their differing capacities, for each CBP, we create a clone with 24K entries, comprising 6 PHTs with 4K entries each, while maintaining similar PHR update and PHT index/tag functions. This allows us to compare the performance solely attributable to the hash functions. We conduct the same experiments using the 24K CBPs. Results show that 24K Firestorm CBP produces only 1\% less MPKI then 24K Oryon and 24K Skylake.

\subsubsection{Conclusion}

We can perceive that the capacity of the CBP significantly contributes to branch prediction accuracy. The CBP of Intel Skylake has the smallest capacity and the highest MPKI. The capacities of Firestorm and Oryon are nearly identical, and their accuracies are also similar. When these CBPs are aligned to the same capacity of 24K entries, the MPKI values become much closer.

\subsection{Comparing PHT Organizations}

In addition to capacity, we compare the design differences in the organization of PHT. For the PHT tag function, Intel, Apple, and Qualcomm appear to agree on PHR folding and XOR-ing PC bits. However, for the PHT index function, Apple and Qualcomm diverge again by using two or three bits XOR-ed instead of folding PHR. Previously we demonstrate that the hash functions contribute less to branch prediction accuracy than capacity, indicating that the differences are primarily due to distinct methods of improving timing.

Regarding PHT partitioning, Half \& Half reveals that Intel CBP is partitioned by one PC bit: PC[5] since Skylake and PC[4] on older microarchitectures \cite{halfhalf}. Similarly, Apple and Qualcomm CBPs are also partitioned by one PC bit: PC[6], according to our CBP models. This is not coincidental, as it enables a significant improvement: the CBP can easily predict the direction of two adjacent conditional branches using the same PHR. They reside in the same prediction block, ensuring that their PC[6] bit is always equal on Firestorm and Qualcomm (likewise on Intel). Consequently, their index bits are identical, allowing the CBP to extract all ways from one set and compare against two tags simultaneously. This approach is both easier and more cost-effective to implement compared to having two read ports for the backing SRAM arrays.

\section{Conclusions}

In this paper, we present, for the first time, the CBP structure of Apple Firestorm and Qualcomm Oryon processors, enabling the construction of a branch predictor model. We introduce a novel reverse engineering pipeline with fewer assumptions, making it more applicable to a broader range of microarchitectures. Subsequently, we identify the \textbf{Scatter} and \textbf{Annihilation} effects and discuss how to utilize the CBP model to guide optimization. Finally, we compare the MPKI of different CBP designs running SPEC CPU 2017 and Geekbench 5 benchmarks, and offer the following insights on CBP design:

\begin{enumerate}
  \item The PHR footprint should include more bits of both branch and target addresses, especially low address bits, to avoid \textbf{Annihilation} effect.
  \item In CBP design, capacity is a critical factor for prediction accuracy. The PHR update and PHT index/tag functions have a smaller impact and allow for more timing optimization.
  \item Predicting two adjacent conditional branches per cycle can be implemented by partitioning PHT by one PC bit and matching two ways simultaneously, without requiring an additional read port for SRAM.
\end{enumerate}

\bibliographystyle{plain}
\bibliography{references}

\begin{thebibliography}{10}

\bibitem{cbp4}
Championship branch prediction (cbp-4), 2014.
\newblock Accessed on Oct, 2024.

\bibitem{cbp5}
Championship branch prediction (cbp-5), 2016.
\newblock Accessed on Oct, 2024.

\bibitem{zen2}
AMD.
\newblock Software optimization guide for amd epyc™ 7002 processors, 2020.
\newblock Accessed on Oct, 2024.

\bibitem{zen3}
AMD.
\newblock Software optimization guide for amd epyc™ 7003 processors, 2020.
\newblock Accessed on Oct, 2024.

\bibitem{zen1}
AMD.
\newblock Software optimization guide for amd family 17h processors, 2021.
\newblock Accessed on Oct, 2024.

\bibitem{zen4}
AMD.
\newblock Software optimization guide for the amd zen4 microarchitecture, 2023.
\newblock Accessed on Oct, 2024.

\bibitem{zen5}
AMD.
\newblock Software optimization guide for the amd zen5 microarchitecture, 2024.
\newblock Accessed on Oct, 2024.

\bibitem{armv8}
ARM.
\newblock Arm architecture reference manual armv8 for armv8-a architecture
  profile, 2024.
\newblock Accessed on Oct, 2024.

\bibitem{spec2017}
James Bucek, Klaus-Dieter Lange, and J\'{o}akim v.~Kistowski.
\newblock Spec cpu2017: Next-generation compute benchmark.
\newblock In {\em Companion of the 2018 ACM/SPEC International Conference on
  Performance Engineering}, ICPE '18, page 41–42, New York, NY, USA, 2018.
  Association for Computing Machinery.

\bibitem{zen2analysis}
Ian Cutress.
\newblock Amd zen 2 microarchitecture analysis: Ryzen 3000 and epyc rome, 2019.
\newblock Accessed on Oct, 2024.

\bibitem{champsim}
Nathan Gober, Gino Chacon, Lei Wang, Paul~V. Gratz, Daniel~A. Jimenez, Elvira
  Teran, Seth Pugsley, and Jinchun Kim.
\newblock {The Championship Simulator: Architectural Simulation for Education
  and Competition}.
\newblock {\em arXiv preprint arXiv:2210.14324}, 2022.

\bibitem{exynos}
Brian Grayson, Jeff Rupley, Gerald~Zuraski Zuraski, Eric Quinnell, Daniel~A
  Jim{\'e}nez, Tarun Nakra, Paul Kitchin, Ryan Hensley, Edward Brekelbaum,
  Vikas Sinha, et~al.
\newblock Evolution of the samsung exynos cpu microarchitecture.
\newblock In {\em 2020 ACM/IEEE 47th Annual International Symposium on Computer
  Architecture (ISCA)}, pages 40--51. IEEE, 2020.

\bibitem{geekbench5}
Primate Labs.
\newblock https://www.geekbench.com/blog/2019/09/geekbench-5/, 2019.
\newblock Accessed on Oct, 2024.

\bibitem{indirector}
Luyi Li, Hosein Yavarzadeh, and Dean Tullsen.
\newblock Indirector:$\{$High-Precision$\}$ branch target injection attacks
  exploiting the indirect branch predictor.
\newblock In {\em 33rd USENIX Security Symposium (USENIX Security 24)}, pages
  2137--2154, 2024.

\bibitem{combining}
S~McFarling.
\newblock Combining branch predictors.
\newblock Technical report, Citeseer, 1993.

\bibitem{phr}
Ravi Nair.
\newblock Dynamic path-based branch correlation.
\newblock In {\em Proceedings of the 28th annual international symposium on
  Microarchitecture}, pages 15--23. IEEE, 1995.

\bibitem{btb}
Arthur Perais and Rami Sheikh.
\newblock Branch target buffer organizations.
\newblock In {\em Proceedings of the 56th Annual IEEE/ACM International
  Symposium on Microarchitecture}, pages 240--253, 2023.

\bibitem{simpoint}
Erez Perelman, Greg Hamerly, Michael Van~Biesbrouck, Timothy Sherwood, and Brad
  Calder.
\newblock Using simpoint for accurate and efficient simulation.
\newblock {\em ACM SIGMETRICS Performance Evaluation Review}, 31(1):318--319,
  2003.

\bibitem{ogehl2}
Andre Seznec.
\newblock The o-gehl branch predictor.
\newblock {\em The 1st JILP Championship Branch Prediction Competition
  (CBP-1)}, 2004.

\bibitem{ogehl}
Andre Seznec.
\newblock Analysis of the o-geometric history length branch predictor.
\newblock In {\em 32nd International Symposium on Computer Architecture
  (ISCA'05)}, pages 394--405, 2005.

\bibitem{tage}
Andr{\'e} Seznec and Pierre Michaud.
\newblock A case for (partially) tagged geometric history length branch
  prediction.
\newblock {\em The Journal of Instruction-Level Parallelism}, 8:23, 2006.

\bibitem{study}
James~E Smith.
\newblock A study of branch prediction strategies.
\newblock In {\em 25 years of the international symposia on Computer
  architecture (selected papers)}, pages 202--215, 1998.

\bibitem{reduction}
Robert~G Wedig and Marc~A Rose.
\newblock The reduction of branch instruction execution overhead using
  structured control flow.
\newblock In {\em Proceedings of the 11th annual international symposium on
  Computer architecture}, pages 119--125, 1984.

\bibitem{hotchips}
Gerard Williams.
\newblock Snapdragon x elite qualcomm oryon cpu: Design \& architecture
  overview.
\newblock In {\em HCS}, 2024.

\bibitem{trustzonetunnel}
Tianhong Xu, Aidong~Adam Ding, and Yunsi Fei.
\newblock Trustzonetunnel: A cross-world pattern history table-based
  microarchitectural side-channel attack.
\newblock In {\em 2024 IEEE International Symposium on Hardware Oriented
  Security and Trust (HOST)}, pages 01--11. IEEE, 2024.

\bibitem{halfhalf}
Hosein Yavarzadeh, Mohammadkazem Taram, Shravan Narayan, Deian Stefan, and Dean
  Tullsen.
\newblock Half\&half: Demystifying intel’s directional branch predictors for
  fast, secure partitioned execution.
\newblock In {\em 2023 IEEE Symposium on Security and Privacy (SP)}, pages
  1220--1237. IEEE, 2023.

\bibitem{ghr}
Tse-Yu Yeh and Yale~N Patt.
\newblock Alternative implementations of two-level adaptive branch prediction.
\newblock {\em ACM SIGARCH Computer Architecture News}, 20(2):124--134, 1992.

\bibitem{spectre}
Google~Project Zero.
\newblock Reading privileged memory with a side-channel, 2018.
\newblock Accessed on Oct, 2024.

\end{thebibliography}

\end{document}